\documentclass{article}

\usepackage{microtype}
\usepackage{graphicx}
\usepackage{subfigure}
\usepackage{booktabs} 

\usepackage{hyperref}

\usepackage[accepted]{icml2025}

\usepackage{amsmath}
\usepackage{amssymb}
\usepackage{mathtools}
\usepackage{amsthm}

\usepackage[capitalize,noabbrev]{cleveref}

\theoremstyle{plain}

\theoremstyle{definition}

\theoremstyle{remark}

\usepackage[textsize=tiny]{todonotes}

\icmltitlerunning{Self-supervised learning predicts plant growth trajectories from multi-modal industrial greenhouse data}

\begin{document}

\twocolumn[
\icmltitle{Self-supervised learning predicts plant growth trajectories from \\ multi-modal industrial greenhouse data}




\begin{icmlauthorlist}
\icmlauthor{Adam J Riesselman}{hh}
\icmlauthor{Evan M Cofer}{hh}
\icmlauthor{Therese LaRue}{hh}
\icmlauthor{Wim Meeussen}{hh}
\end{icmlauthorlist}

\icmlaffiliation{hh}{Hippo Harvest}

\icmlcorrespondingauthor{Adam J Riesselman}{adam@hippoharvest.com}
\icmlcorrespondingauthor{Wim Meeussen}{wim@hippoharvest.com}

\icmlkeywords{Machine Learning, ICML}

\vskip 0.3in
]



\printAffiliationsAndNotice{} 

\begin{abstract}
Quantifying organism-level phenotypes, such as growth dynamics and biomass accumulation, is fundamental to understanding agronomic traits and optimizing crop production. However, quality growing data of plants at scale is difficult to generate. Here we use a mobile robotic platform to capture high-resolution environmental sensing and phenotyping measurements of a large-scale hydroponic leafy greens system. We describe a self-supervised modeling approach to build a map from observed growing data to the entire plant growth trajectory. We demonstrate our approach by forecasting future plant height and harvest mass of crops in this system. This approach represents a significant advance in combining robotic automation and machine learning, as well as providing actionable insights for agronomic research and operational efficiency. 

\end{abstract}

\section{Introduction}
\label{introduction}

Understanding how plants grow is crucial for predicting biomass production and maturity; these important agronomic traits are directly linked to crop yield. However, reliably estimating these traits is challenging because plant growth is sensitive to the environment, and varies across species and cultivars. Robustly capturing the relationships between genotype, environment, and organismal phenotype demands extensive data.
\begin{figure}[t]
  \centering
  \includegraphics[width=\linewidth]{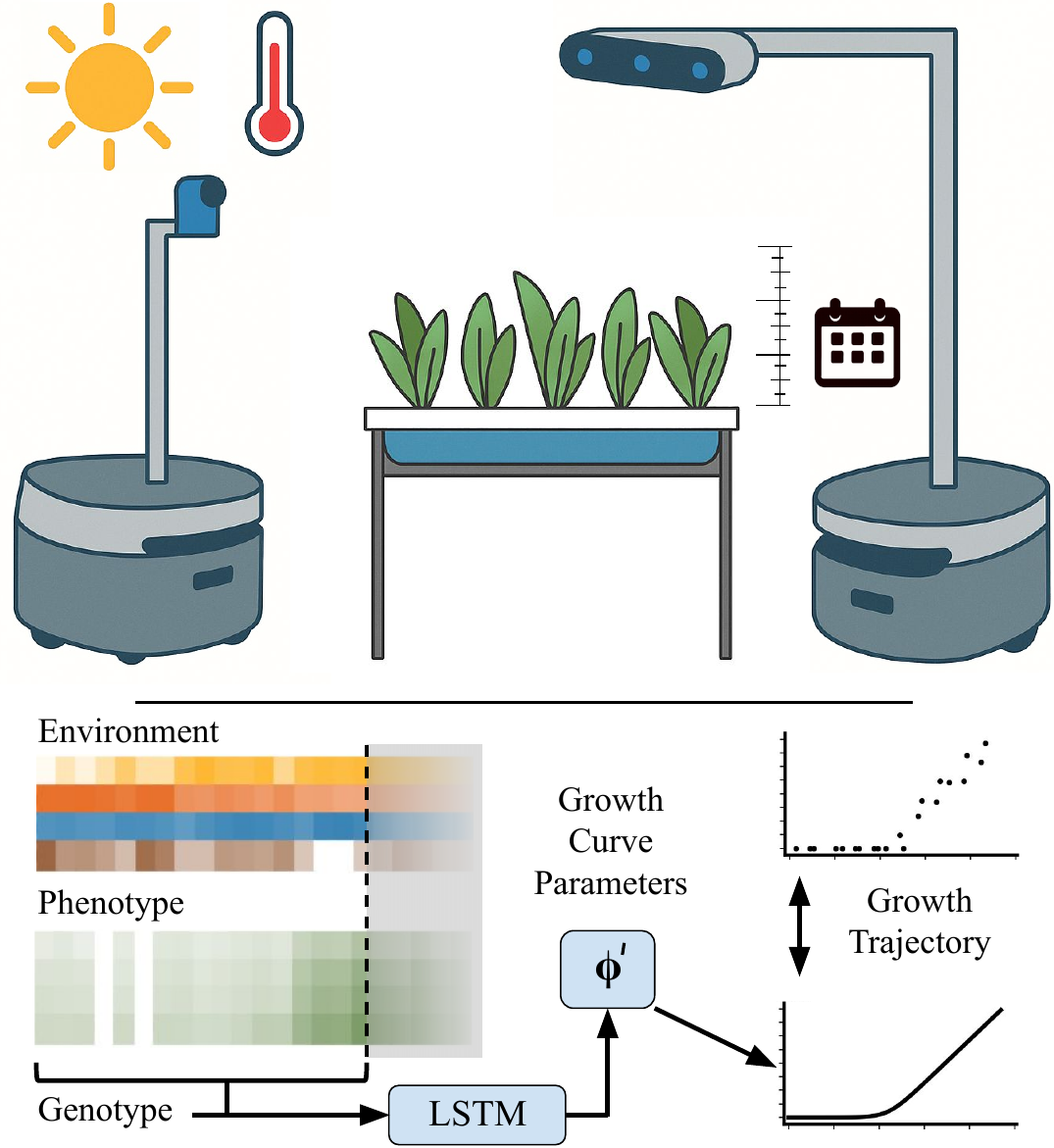}
  \caption{Top: Robots and sensors collect data of leafy greens grown hydroponically: left logs environment (temperature, light, humidity), and right captures phenotype (height over time). Bottom: The entire trajectory of growth is generated using currently observed environment, phenotype, and genotype data.}
  \label{fig:overview}
\end{figure}

Traditionally, efforts to build such predictive models have been hindered by the cost and labor involved in collecting detailed phenotypic and environmental information. In this study, we deploy an automated robotic system managing a large-scale hydroponic production facility growing lettuce and spinach varieties. These robots systematically gather high-dimensional environmental and phenotypic data across thousands of growing trays daily.

Leveraging this comprehensive dataset, we develop HINTS (Harvest Intelligence from Neural Time Series), a generative model that links environmental and phenotypic inputs to key outcomes like growth rates and harvest mass. Our approach uniquely benefits from self-supervision, where the natural physical growth process provides the supervisory signal. Models trained on this data can generate growth parameters, which can be used to predict plant height and weight at harvest. This approach can fully leverage automated robotic phenotyping and environmental sensing to understand plant biology.

\section{Related Work}
A standard approach to estimating plant maturity is growing degree days: This heuristic models plant growth as the accumulation of heat each day throughout the life cycle, with species or variety specific temperature thresholds \cite{de1735observation,miller2001using}. These approaches have been further refined by incorporating additional constraints, like plant stress \cite{hatfield2011climate}, and environmental data \cite{brechner1996hydroponic}. In addition, statistical models of plant growth \cite{hunt1982plant} and machine learning methods \cite{sharmin2025review} can improve estimates agricultural productivity.

The development of accurate models require substantial amounts of quality data. High-throughput phenotyping facilities automate this data collection process \cite{abebe2023image, atefi2021robotic} using a combination of computer vision and sensor data to estimate plant growth and maturity \cite{li2021high,xu2022smart}. However, these facilities are expensive to build and maintain, and are not integrated into large-scale commercial operations.

Extracting useful features to represent biology from vast datasets is also a challenge \cite{wang2023scientific}. Self-supervised learning has been frequently employed to build biological insights without extensive labeled data \cite{liu2021self, lopez2018deep, notin2024machine, lu2019learning, brixi2025genome}, and has been incorporated into industrial phenotyping efforts in biomedical research \cite{fay2023rxrx3, sivanandan2023pooled}. These computational approaches, when paired with robotic data acquisition, make them scalable biology learners.

\section{Methods}

\subsection{Data collection}

Lettuce and spinach are cultivated hydroponically in an approximately 1 m$^2$ growing tray across an 3000 m$^2$ sized facility (Figure \ref{fig:overview}). Hydroponic water volume is managed by custom-outfitted freight robots \cite{meeussen2021grow}: one set of robots determines water volume in each tray, while another set delivers precise amounts of water and nutrients to sustain growth.

Additional robots capture environmental and phenotypic data \cite{meeussen2024grow}. One robot is outfitted with a photosynthetically active radiation (par) meter \cite{apogee_sq100x}, measuring light intensity across the growspace (Figure \ref{fig:overview}). Using additional machine learning techniques \cite{riesselman2023high}, we impute sunlight and LED intensity at 15 minute intervals for each growing tray throughout its lifecycle. Ambient air temperature and humidity are also recorded at regular intervals across the growspace.

Phenotypic information is gathered using Intel RealSense D455 depth cameras \cite{intel_d455} mounted on a mobile robotic platform that traverses the growing area daily (Figure \ref{fig:overview}). The images are automatically cropped to the boundaries of individual growing trays, treating each as a distinct sample. From the depth information, we extract plant height measurements, summarized as the median height in centimeters across the growing tray. Additionally, we featurize the height distribution across the tray by calculating 10 height deciles to capture the variability of plant canopy composition.

Paired environment and height measurement data are recorded daily for each growing tray. Environmental data ($E$) are summarized by calculating daily minimum, mean, and maximum values for air temperature and humidity. We also compute the daily light integral (DLI) separately for sunlight and LED lighting systems. These environmental features are paired with the corresponding daily height measurements ($H$). From transplant $0$ to harvest $D$, we organize these observations as sequences $E = \{e_0, e_1, ..., e_D\}$ and $H = \{h_0, h_1, ..., h_D\}$. Finally, the mass of harvested leaf material ($m_D$) is recorded in grams by weighing the growing tray before and after harvest. This results in $657663$ growing days of environmental data, $639352$ phenotypic data points, and $28410$ total harvested growing trays.

\subsection{Height growth curve}

We parameterize plant growth trajectories with a differentiable growth curve by summarizing observed plant height with two key parameters (Figure \ref{fig:growthcurve}): seedling lag ($\beta_{\text{lag}}$) and growth rate ($\beta_{\text{gr}}$):

\begin{equation}
\hat{h}_d = \beta_{\text{gr}} \cdot \text{softplus}(\text{age}_d - \beta_{\text{lag}}) = \beta_{\text{gr}} \cdot \ln(1 + e^{(\text{age}_d - \beta_{\text{lag}})})
\end{equation}

Where $d \in \{0, 1, ..., D\}$ indexes the measurement day, $\text{age}_d$ is the number of days after transplant, $\beta_{\text{lag}}$ represents the initial lag period during which the plant establishes itself before detectable vertical growth begins, and $\beta_{\text{gr}}$ determines the rate of height increase after this establishment phase. While height accumulation of lettuce and spinach does ultimately plateau, we harvest it before this point.

We regress out the effect of sun elevation on observed plant height, as it is a significant covariate in observed height measurements:

\begin{equation}
h_d^{\text{corrected}} = h_d - \beta \cdot \mathbb{1}_{\text{sun up}} \cdot h_d
\end{equation}

Where $\beta$ is a learned parameter and $\mathbb{1}_{\text{sun up}}$ is an indicator function that equals 1 when the sun is above the horizon.

\subsection{Canopy mass}

\begin{figure}[t]
  \centering
  \includegraphics[width=\linewidth]{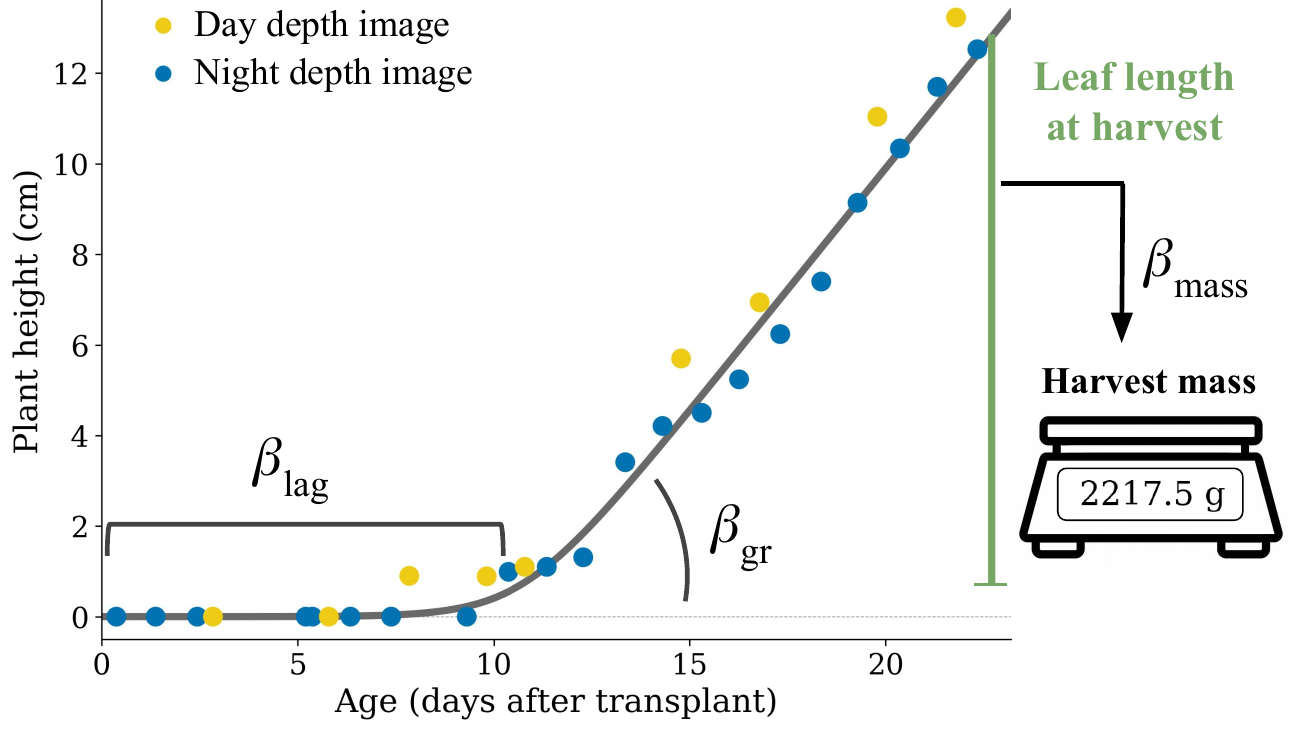}
  \caption{Phenotypic observations and growth parameterization of a harvested growing tray. $\beta_{lag}$ and $\beta_{gr}$ map plant age ($age_{d}$ to estimated height. Harvest mass $m_D$ is estimated from the leaf length at harvest and $\beta_{mass}$.}
  \label{fig:growthcurve}
\end{figure}

To estimate canopy mass, we learn a relative canopy density parameter that relates plant height to mass (Figure \ref{fig:growthcurve}). Here we assume the density is the same at any point in its lifecycle, and that canopy composition is isotropic. Harvested leaf length $l_D$ is calculated by subtracting the cut height $c$ from the total plant height $h_D$:

\begin{equation}
l_D = h_D - c
\end{equation}

Cut height $c$ is set per growing tray for desired leaf product specifications. We predict harvest mass $m_D$ as the product of $l_D$ and a learned density parameter $\beta_{\text{mass}}$:

\begin{equation}
\hat{m}_D = l_D \cdot \beta_{\text{mass}}
\end{equation}

This parameterization uses leaf length as a proxy for the volume of harvestable product, with the density $\beta_{\text{mass}}$ only observed at harvest.

\subsection{Parameter estimation}

We can directly estimate these parameters individually for each growing tray when harvested using the collected plant heights $H_{0:D}$, ages of height observations $A_{0:D}$ and harvest mass $m_D$:

\begin{equation}
P(H_{0:D},m_D|A_{0:D},\phi)
\end{equation}

where $\phi = \{\beta_{\text{lag}}, \beta_{\text{gr}}, \beta_{\text{mass}}\}$ and $\phi \geq 0$.

\subsection{Deep generative model}

While direct parameter estimation at harvest time is straightforward, our goal is to predict these parameters in the middle of a plant's lifecycle, which in turn can be used to project future growth. To accomplish this, we model the relationship between observed data and growth parameters as a nonlinear function $f$ with parameters $\theta$:

\begin{equation}
f_\theta(g, E_{0:d}, \tilde{H}_{0:d}) = [\beta_{\text{lag},d}, \beta_{\text{gr},d}, \beta_{\text{mass},d}] = \phi'_{d}
\end{equation}

where $g$ represents genotype information, $E_{0:d} = \{e_0, e_1, ..., e_d\}$ represents all environmental observations up to day $d$, and $\tilde{H}_{0:d} = \{\tilde{h}_0, \tilde{h}_1, ..., \tilde{h}_d\}$ represents all height feature observations up to day $d$. We parameterize the function $f_\theta$ using a unidirectional LSTM \cite{hochreiter1997long, merrill2024illusion}. All generated growth parameters $\phi'$ are constrained to be positive through a softplus transformation.
At each time step $d$, the model takes all environmental $E_{0:d}$ and height feature $\tilde{H}_{0:d}$ data observed from the beginning of the cycle up to that point; missing height data are masked with zeros (Figure 1). Growth parameters $\phi'$ are generated at each step, $[\phi'_{0}...\phi'_d]$, conditioned on previously observed environmental, phenotypic, and genetic data.

These parameters can then be used to transform plant age (days after transplant) at time $t$ into predicted height $\hat{h}_{dt}$:

\begin{equation}
\hat{h}_{dt} = \beta_{\text{gr},d} \cdot \text{softplus}(\text{age}_t - \beta_{\text{lag},d})
\end{equation}

To predict leaf mass at harvest, we employ a more flexible vector-based approach to model variable leaf densities along different sections of the plant:

\begin{equation}
\hat{m}_D = \sum_{j=1}^{k} (\boldsymbol{\beta}_{\text{mass},d,j} \cdot \mathbf{I}_j)
\end{equation}

Where $\boldsymbol{\beta}_{\text{mass},d} \in \mathbb{R}^k$ is a vector of $k$ canopy density parameters output by the model at day $d$, and $\mathbf{I} \in \{0,1\}^k$ is a vector of indicator variables representing the median harvested leaf length per tray $l_{dt}$. This formulation allows realization of variable canopy density across different sections of the leaf but allow for variable cut height $c$, which is not input directly into the model, to prevent overfitting.

Collectively, these components define a model that estimates the conditional probability distribution:

\begin{equation}
P(H_{0:D},m_D|E_{0:d},\tilde{H}_{0:d},A_{0:d},g;\theta)
\end{equation}

where $H_{0:D}$ represents median plant height and $\tilde{H}_{0:d}$ represents height features (height deciles), respectively, for the entire cycle length, from time $0$ to harvest day $D$. The model is end-to-end differentiable, optimizing LSTM parameters while also enforcing structure and interpretability in phenotype characterization through $\phi'$.

\subsection{Objective function and training}

We implemented our model using a 3 layer residual LSTM architecture in Pytorch following \cite{beck2024xlstm}. The LSTM consists of 3 layers with a dropout rate of 0.05. Input features from environment, height observations, and genotype are embedded into a common 128-dimensional representation space (Appendix).

The overall objective function for our model combines losses for plant height and harvest mass predictions with regularization terms that provide disambiguation when limited observations are available:

\begin{equation}
\mathcal{L} = \mathcal{L}_{\text{height}} + \mathcal{L}_{\text{mass}} + \mathcal{L}_{\text{prior}}
\end{equation}

Where $\mathcal{L}_{\text{height}}$ and $\mathcal{L}_{\text{mass}}$ are Pseudo-Huber losses \cite{charbonnier1997deterministic} for plant height and harvest mass predictions respectively, and $\mathcal{L}_{\text{prior}}$ encompasses four regularization terms. We use Pseudo-Huber loss due to its robustness against outliers in our robot-collected data. The complete mathematical formulation and hyperparameters of all six individual loss terms are provided in the Appendix.

We trained the model using the AdamW optimizer \cite{loshchilov2017decoupled} with a learning rate of 0.001 and early stopping with a patience of 10 epochs.

\section{Experiments}

A key operational challenge in hydroponic production is accurately forecasting harvest mass with sufficient lead time to meet customer fulfillments.
In our system, we need to predict the harvest mass 5 days in advance of the actual harvest date to ensure accurate order fulfillment.

For each growing tray in the growspace, we collate all environmental and phenotypic data available up until the planning date, and use these data in conjunction with the genotype features to predict plant growth and relative canopy density parameters. We then project height and harvest mass 5 days into the future.
As a baseline, we compare the HINTS-predicted harvest height and mass estimates to those from rolling N-day genotype-specific historic parameter averages for $\beta_\text{lag}$, $\beta_\text{gr}$, and $m_{D}$ for $N\in\left\{10,30,90\right\}$.
We evaluated the performance of HINTS and each baseline using $1989$ lettuce growing trays harvested in April of 2025.

\begin{figure}[htbp]
  \centering
  \includegraphics[width=0.9\linewidth, trim=0 5 0 0, clip]{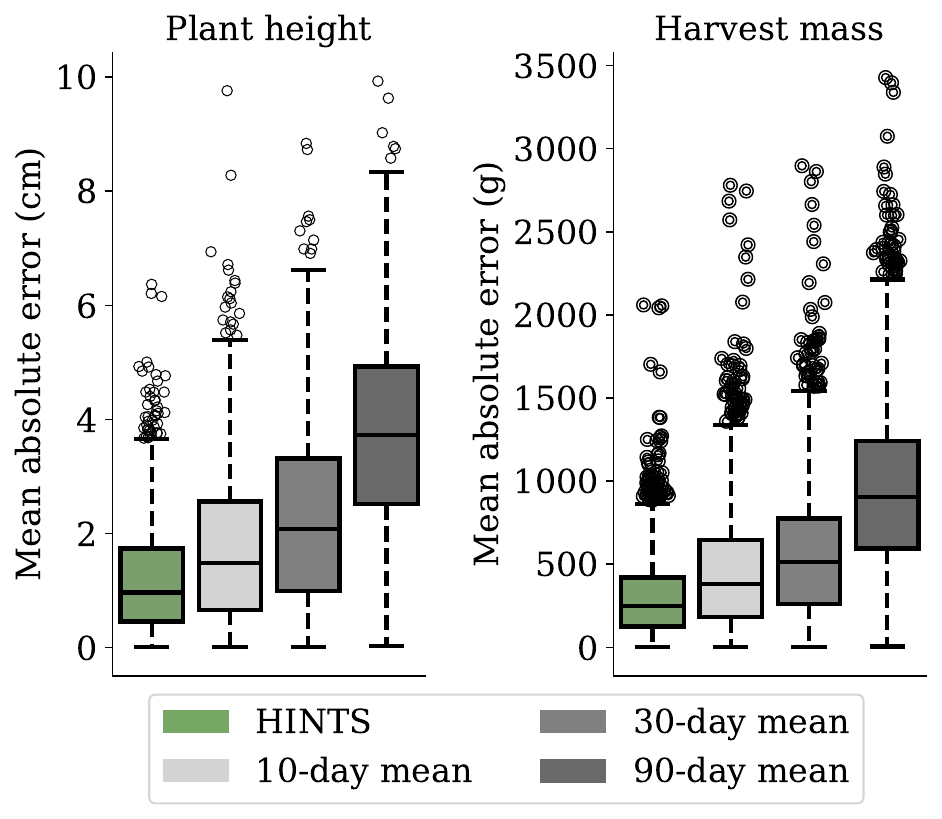}\caption{Absolute error comparison of HINTS and a baseline using $N$-day rolling average parameters. Performance is evaluated using $1,989$ lettuce growing trays }\label{fig:performance}
\end{figure}

As shown in Figure \ref{fig:performance}, HINTS outperforms each of the $N$-day rolling average parameter baselines.
Specifically, HINTS outperforms the 10-, 30-, and 90-day baseline predictions for plant height at harvest by $29.92\%$, $45.92\%$, and $67.37\%$ mean absolute error.
It similarly outperformed the baseline predictions for harvest mass (g) by $33.93\%$, $45.99\%$, and $67.98\%$ respectively.
This demonstrates HINTS's ability to successfully recover growth dynamics and harvest characteristics by combining partial growth trajectories with environmental features.

\section{Conclusion}

We present a generative, self-supervised model, HINTS, that accurately forecasts plant height and harvest mass in a large-scale hydroponic system by leveraging environmental, phenotypic, and genotypic data.
This approach outperforms conventional heuristic baselines using rolling averages for parameters, achieving at least a $29.92\%$ improvement in harvest height prediction, and $33.93\%$ harvest mass prediction.
Accurately predicting these phenotypes enables successful production planning while simultaneously minimizing wasted resources and food.

Importantly, HINTS maintains interpretability through biologically meaningful parameters (e.g. $\beta_\text{lag}$, $\beta_\text{gr}$), allowing operators to reason about predictions and optimize growing conditions proactively.
These findings demonstrate the utility of combining automated phenotyping with structured deep learning for controlled environment agriculture.

\section*{Impact Statement}
This work addresses critical challenges in food security and climate change by improving our ability to predict and optimize crop yields in controlled environment agriculture. Methods herein enable more accurate predictions of plant growth and resource allocation as well as reduce food waste.

\section*{Acknowledgements}
We would like to thank the entire Hippo Harvest team for their contributions to this work, particularly Eitan Marder-Eppstein, Chelsea Wirth, Kevin Hannegan, Zack Lewis, Lucas Becker-Lipton, and Teka Nicholas.

\bibliography{GenBioWorkshop2025_Final}
\bibliographystyle{icml2025}

\newpage
\appendix
\onecolumn
\section{Appendix}

\subsection{Growspace facility description}

The growspace consists of greenhouses outfitted with LEDs, shades, and climate control systems. Growing trays with plants are organized into a grid and occupy as much growspace real estate as possible. Robots travel beneath the trays: gaps between columns give space for robot components to be mounted on a frame above the plants. This configuration is built for in situ robot operations, like imaging, weighing, or environmental sensing.

\begin{figure}[h]
  \centering
  \includegraphics[width=0.9\textwidth]{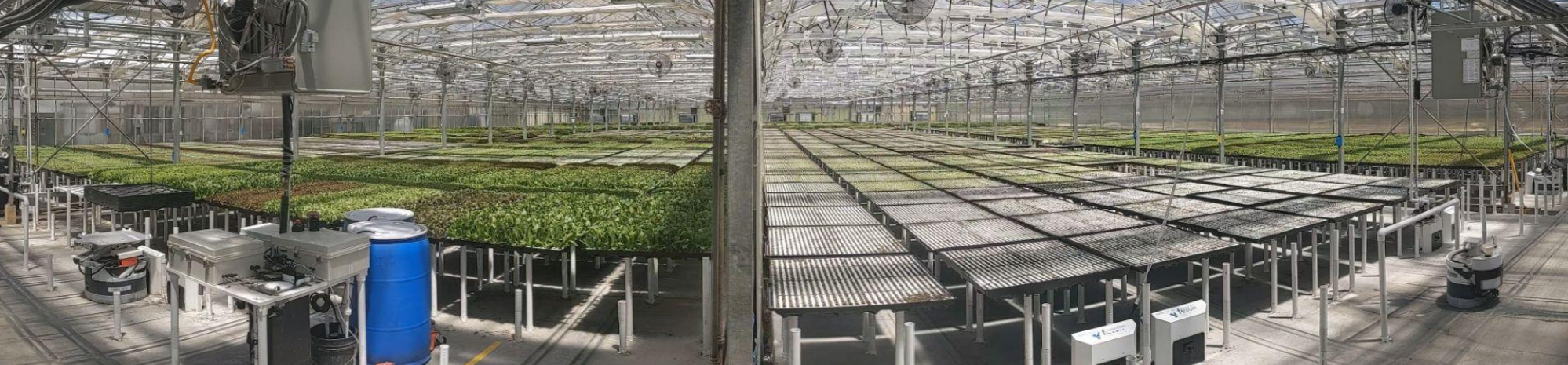}
  \caption{A view of a hydroponic leafy greens growspace.}
  \label{fig:facility}
\end{figure}

\subsection{Detailed Loss Function Formulation}

The model's objective function consists of six distinct terms, with loss weights applied to the prior terms as follows:

\begin{equation}
\mathcal{L} = \mathcal{L}_{\text{height}} + \mathcal{L}_{\text{mass}} + 0.01 \cdot \mathcal{L}_{\text{embed}} + 0.1 \cdot \mathcal{L}_{\text{density}} + 0.01 \cdot \mathcal{L}_{\text{lag}} + 0.1 \cdot \mathcal{L}_{\text{growth}}
\end{equation}

The individual terms are defined as follows:

\begin{enumerate}
\item $\mathcal{L}_{\text{height}}$ is the Pseudo-Huber loss for plant height predictions:
   \begin{equation}
   \mathcal{L}_{\text{height}} = \sum_{d} \mathcal{H}_{\delta_h}(h_d^{\text{corrected}} - \hat{h}_d)
   \end{equation}
   with $\delta_h = 3$ cm (transition point from quadratic to linear loss)

\item $\mathcal{L}_{\text{mass}}$ is the Pseudo-Huber loss for harvest mass predictions:
   \begin{equation}
   \mathcal{L}_{\text{mass}} = \mathcal{H}_{\delta_m}(m_D - \hat{m}_D)
   \end{equation}
   with $\delta_m = 2$ kg

The Pseudo-Huber loss $\mathcal{H}_\delta$ is defined as:
\begin{equation}
\mathcal{H}_\delta(x) = \delta^2 \left(\sqrt{1 + \left(\frac{x}{\delta}\right)^2} - 1\right)
\end{equation}

This differentiable approximation of the Huber loss provides robustness against outliers while maintaining differentiability at all points. We use it for both height and mass predictions due to occasional spurious measurements in our robot-collected data.

\item $\mathcal{L}_{\text{lag}}$ is the L1 prior loss on the seedling lag parameter:
   \begin{equation}
   \mathcal{L}_{\text{lag}} = |\beta_{\text{lag}} - 12|
   \end{equation}
   where $\beta_{\text{lag}}$ represents the seedling lag parameter with prior mean 12

\item $\mathcal{L}_{\text{growth}}$ is the L1 prior loss on the growth rate parameter:
   \begin{equation}
   \mathcal{L}_{\text{growth}} = |\beta_{\text{gr}} - (-0.5)|
   \end{equation}
   where $\beta_{\text{gr}}$ represents the growth rate parameter with prior mean -0.5

\item $\mathcal{L}_{\text{density}}$ is the L1 prior loss on the canopy density parameter:
   \begin{equation}
   \mathcal{L}_{\text{density}} = |\beta_{\text{mass}} - (-2)|
   \end{equation}
   where $\beta_{\text{mass}}$ represents the canopy density parameter with prior mean -2

\item $\mathcal{L}_{\text{embed}}$ is the L1 regularization on the genotype embedding weights that project the genotype embedding to the hidden unit size:
   \begin{equation}
   \mathcal{L}_{\text{embed}} = \frac{1}{|\mathbf{W}|}\sum_{i,j}|W_{i,j}|
   \end{equation}
   where $\mathbf{W}$ is the weight matrix of the genotype embedding layer
\end{enumerate}

In particular, priors on generated parameters $\beta_{\text{gr}}$ and $\beta_{\text{lag}}$ provide disambiguation when limited height observations are available, or all zeros.

\begin{table}[h]
\centering
\begin{tabular}{|c|c|c|l|}
\hline
\textbf{Parameter} & \textbf{Weight Coefficient} & \textbf{Prior Mean} & \textbf{Description} \\ \hline
Embedding prior & 0.01 & 0 & Genotype embedding regularization \\ \hline
Canopy density prior & 0.1 & -2 & Weight for the canopy density parameter \\ \hline
Seedling lag prior & 0.01 & 12 & Weight for the seedling lag parameter \\ \hline
Growth rate prior & 0.1 & -0.5 & Weight for the growth rate parameter \\ \hline
\end{tabular}
\caption{Weight coefficients for the L1 priors in the objective function}
\label{tab:weight_coeffs}
\end{table}

\subsection{Genotype Encoding}

Genotype information is included into the model by MD5 hash-based encoding, where each plant variety string is hashed to produce a stable 16-byte representation. This vector is then normalized to the range [0,1] before being projected to the model's embedding space.


\end{document}